\documentclass[aps, prl, superscriptaddress, twocolumn, 10pt]{revtex4-1}

\usepackage{amsmath}	% AMS Math Package
\usepackage{amsthm}		% Theorem Formatting
\usepackage{amssymb}	% Math symbols such as \mathbb
\usepackage{array}
\usepackage{datetime}
\usepackage{graphicx}
\usepackage{color}
\usepackage{verbatim}
\usepackage{bm}
\usepackage{braket}
\usepackage{natbib}
\usepackage{cancel}
\usepackage{xcolor}
\usepackage{slashed}
\usepackage{braket}
\usepackage{amsfonts} 
\smallskipamount
\usepackage{feynmp}
\usepackage{letltxmacro} % closed roots
\usepackage[colorlinks=true, citecolor=midblue, linkcolor=midblue, urlcolor=midblue]{hyperref}

\usepackage{setspace}
\usepackage[normalem]{ulem}

\usepackage{tikz,xcolor}
\definecolor{lime}{HTML}{A6CE39}
\DeclareRobustCommand{\orcidicon}{
	\begin{tikzpicture}
	\draw[lime, fill=lime] (0,0) 
	circle [radius=0.16] 
	node[white] {{\fontfamily{qag}\selectfont \tiny ID}};
	\draw[white, fill=white] (-0.0625,0.095) 
	circle [radius=0.007];
	\end{tikzpicture}
	\hspace{-2mm}
}
\foreach \x in {A, ..., Z}{
	\expandafter\xdef\csname orcid\x\endcsname{\noexpand\href{https://orcid.org/\csname orcidauthor\x\endcsname}{\noexpand\orcidicon}}
}

\settimeformat{ampmtime}

\definecolor{grey}{rgb}{0.4,0.4,0.4}
\definecolor{dullmagenta}{rgb}{0.4,0,0.4}
\definecolor{darkblue}{rgb}{0,0,0.4}
\definecolor{midblue}{rgb}{0,0,0.5}
\definecolor{midred}{rgb}{0.5,0,0}
\definecolor{orange}{rgb}{1,0.5,0}
\definecolor{lightbrown}{rgb}{0.75,0.5,0.25}
\definecolor{tan}{cmyk}{0.14,0.42,0.56,0}
\definecolor{djunglegreen}{cmyk}{0.99,0,0.52,0}
\definecolor{lightgreen}{rgb}{0,1,0}
\definecolor{olivegreen}{cmyk}{0.64,0,0.95,0.40}
\definecolor{midgreen}{rgb}{0.0,0.675,0.0}
\definecolor{darkgreen}{rgb}{0,0.5,0}

\newcommand{\vs}{\vspace}

\renewcommand{\.}{\hspace{0.5mm}}

\newcommand{\crm}{\ensuremath{\mathrm{c}}}

\newcommand{\Tsf}{\ensuremath{\mathsf{T}}}

\newcommand{\Ibb}{\ensuremath{\mathbb{I}}}

\newcommand{\Nbb}{\ensuremath{\mathbb{N}}}

\newcommand{\Rbb}{\ensuremath{\mathbb{R}}}

\newcommand{\Cbm}{\ensuremath{\bm{C}}}

\newcommand{\Pbm}{\ensuremath{\bm{P}}}

\newcommand{\Wbm}{\ensuremath{\bm{W}}}
\newcommand{\Xbm}{\ensuremath{\bm{X}}}

\newcommand{\kbm}{\ensuremath{\bm{k}}}

\newcommand{\mbm}{\ensuremath{\bm{m}}}

\newcommand{\pbm}{\ensuremath{\bm{p}}}

\newcommand{\xbm}{\ensuremath{\bm{x}}}

\renewcommand{\d}{\ensuremath{\mathrm{d}}}
\newcommand{\vp}{\ensuremath{\vphantom{1^{1^{1^{1^{1}}}}}}}

\settimeformat{ampmtime}

\makeatletter
\let\oldr@@t\r@@t
\def\r@@t#1#2{%
\setbox0=\hbox{$\oldr@@t#1{#2\,}$}\dimen0=\ht0
\advance\dimen0-0.2\ht0
\setbox2=\hbox{\vrule height\ht0 depth -\dimen0}%
{\box0\lower0.4pt\box2}}
\LetLtxMacro{\oldsqrt}{\sqrt}
\renewcommand*{\sqrt}[2][\ ]{\oldsqrt[#1]{#2}}
\makeatother

\newcommand{\FirstAffiliation}{\affiliation{
	Arnold Sommerfeld Center,
	Ludwig-Maximilians-Universit{\"a}t,
	Theresienstra{\ss}e 37,
	80333 M{\"u}nchen,
	Germany}}
 
\newcommand{\SecondAffiliation}{\affiliation{
	Max-Planck-Institut f{\"u}r Physik,
	Boltzmannstr.~8, 
	85748 Garching,
	Germany}}

\newcommand{\ThirdAffiliation}{\affiliation{
	Fakult{\"a}t f{\"u}r Physik, Universit{\"a}t Bielefeld, 
	Postfach 100131, 
	33501 Bielefeld, 
	Germany}}

\begin{document}

%%%%%%%%%%%%%%%%%%%%%%%%%%%%%%%%%%%%%
\title{Aspects of Spatially-Correlated Random Fields:\\ Extreme-Value Statistics and Clustering Properties}

%%%%%%%%%%%%%%%%%%%%%%%%%%%%%%%%%%%%%
\author{Ka Hei Choi\!\orcidA}
\FirstAffiliation

\author{James Creswell}
\FirstAffiliation

\author{Florian K{\"u}hnel\!\orcidC}
\FirstAffiliation
\SecondAffiliation

\author{Dominik J.\ Schwarz\!\orcidD}
\ThirdAffiliation

%%%%%%%%%%%%%%%%%%%%%%%%%%%%%%%%%%%%%
\date{\formatdate{\day }{ \month }{ \year}, \currenttime}

%%%%%%%%%%%%%%%%%%%%%%%%%%%%%%%%%%%%%
\begin{abstract}
\noindent Rare events of large-scale spatially-correlated exponential random fields are studied. The influence of spatial correlations on clustering and non-sphericity is investigated. The size of the performed simulations permits to study beyond-$7.5$-sigma events (one in $10^{13}$). As an application, this allows to resolve individual Hubble patches which fulfil the condition for primordial black hole formation. It is argued that their mass spectrum is drastically altered due to co-collapse of clustered overdensities as well as the mutual threshold-lowering through the latter. Furthermore, the corresponding non-sphericities may imply possibly large changes in the initial black hole spin distribution.\\[-6mm]
\end{abstract}

\maketitle

%%%%%%%%%%%%%%%%%%%%%%%%%%%%%%%%%%%%%
\noindent
{\it Introduction\,---}\;Spatially-correlated random fields are fundamental to numerous physical phenomena, ranging from the microstructure of materials~\cite{Torquato2002} to large-scale patterns in cosmology~\cite{Peebles1980}. Understanding their statistical properties is important for describing for instance phase transitions~\cite{Stanley1971}, pattern formation~\cite{Cross1993} as well as other collective behaviours.

Classical extreme-value theory, pioneered by Fisher \& Tippett~\cite{fisher-tippett-1928} and formalised by Gnedenko~\cite{Gnedenko1943} and Gumbel~\cite{Gumbel1958}, primarily addresses independent and identically distributed random variables. However, many physical systems exhibit significant spatial correlations that invalidate the independence assumption~\cite{Bouchaud1997}. These correlations can lead to substantial deviations from predictions of classical extreme-value statistics, affecting universality classes and scaling behaviours~\cite{Bramwell1998}.

Clustering properties in random fields can be important in many cases, for instance near critical points where fluctuations become long-ranged~\cite{Stauffer1994}. Furthermore, the formation and distribution of clusters have profound implications for percolation processes~\cite{Coniglio1982} and transport phenomena in disordered media~\cite{2002AdPhy..51..187H}. The interplay between extreme events and clustering can, in general, give rise to emergent behaviours that are not evident when these aspects are considered independently (cf.~Ref.~\cite{Sornette2000}).

The statistics of peaks and clustering in Gau{\ss}ian and near-Gau{\ss}ian random fields have been investigated in the literature, especially with cosmological and astrophysical applications~\cite{1985MNRAS.217..805P, 1985ApJ...297...16H, Bardeen:1985tr, Coles:1989su, Matsubara_2020}. Our study extends works such as these by considering  exponentially-distributed discrete random fields. While for (spatially-correlated) Gau{\ss}ian random fields, not only many studies have been made, but, for a class of power spectra, even analytical formul{\ae} have been derived (cf.~Ref.~\cite{adler2010geometry}), this is not the case for spatially-correlated exponential random fields. The latter have become fundamentally important in cosmological contexts, especially for primordial black holes (PBHs)~\cite{ZeldovichNovikov69, Carr:1974nx}, whose probability distribution in many cases has exponential tails (cf.~Refs.~\cite{Biagetti:2018pjj, Ezquiaga:2019ftu, Tada:2021zzj}). For our studies, we have generated data of a much larger size than any previous work, which enables further precision analysis of block maxima estimators for exponential random fields, for which no analytical expressions exist.

In this {\it Letter}, we discuss results of our simulation of spatially-correlated exponential random fields, being the currently largest one. We first focus on the extreme-value statistics emerging through block maxima, and then turn to the study of clustering in dependence of the strength of the spatial correlation. Here, we highlight the various emerging non-spherical shapes. This is then followed by a discussion of the application of our results to the formation of black holes in the early Universe. Finally, we conclude with a discussion of our results and an outlook.

%%%%%%%%%%%%%%%%%%%%%%%%%%%%%%%%%%%%%
\noindent
{\it Simulation of Correlated Random Fields\,---}\vp\;In this work, we have simulated more than $10^{7}$ discrete random fields $F\!: \Rbb^{3} \rightarrow \Rbb$, with each simulation having $N \times N \times N$ pixels for $N = 2^{9}$. The utilised class of power spectra is
\begin{align}
	P( k )
		&\propto
			k^{n}
			\, ,
			\label{eq:Power-Spectrum}
			\\[-6mm]
\intertext{where\vs{-2mm}}
	\big\langle F_{\kbm}\.F_{\pbm} \big\rangle
		&=
			\delta^{3}( \kbm + \pbm )\.P( k )
			\, ,
\end{align}
with $F_{\kbm}$ being the Fourier transform of $F( \xbm )$, $k \equiv | \kbm |$; the spectral index $n$ is taken to be either $0$ (white noise), $-1$, $-2$ or $-3$, for definiteness. In order to address the infrared divergence of $P( k )$ for $n < 0$ at $k = 0$, we imposed an infrared cutoff to remove only the zero mode.

The generation of spatially-correlated exponential random fields $E$ is performed by first generating spatially-correlated Gau{\ss}ian random fields utilising the discrete convolution method outlined in Ref.~\cite{Bertschinger:2001ng}: Let $\mbm$ denote an integer triplet with components $m_{i} \in [ 0 , N -1 ]$ for $i = 1,\.2,\.3$, which specifies the grid positions $\xbm( \mbm ) = ( L/N )\,\mbm$ with a simulation cube of length $L$. Then
\begin{equation}
\label{eq:discrete-fourier}
	F( \mbm )
		=
			\frac{ 1 }{ N^{3} }\.
			\sum_{\bm{\kappa}}\,\exp\!
			\left(
				\imath \.\frac{ 2\pi }{ N }\.
				\bm{\kappa} \cdot \mbm
			\right)
			\mspace{-2mu}
			\sqrt{P( k )}\;
			\xi( \kbm )
			\, ,
\end{equation}
is a Gau{\ss}ian random field of a given power spectrum $P( k )$, the dimensionless wave vector $\bm{\kappa} = \kbm\,L/(2\pi)$ has components $\kappa_{i} \in [\mspace{1mu}- N/2,\.N/2\mspace{1mu})$ for $i = 1,\.2,\.3$, and $\xi( k ) = \exp\!\big[\mspace{1mu} \imath \mspace{1mu}\varphi( \kbm ) \big]$ is a stochastic field, with each randomised phase drawn independently from a uniform distribution, $\varphi( \kbm ) \sim U[\mspace{1mu}-\pi,\.\pi\mspace{1mu}]$.

\begin{figure}[t!] 
	\centering
	\includegraphics[width=0.98\columnwidth]{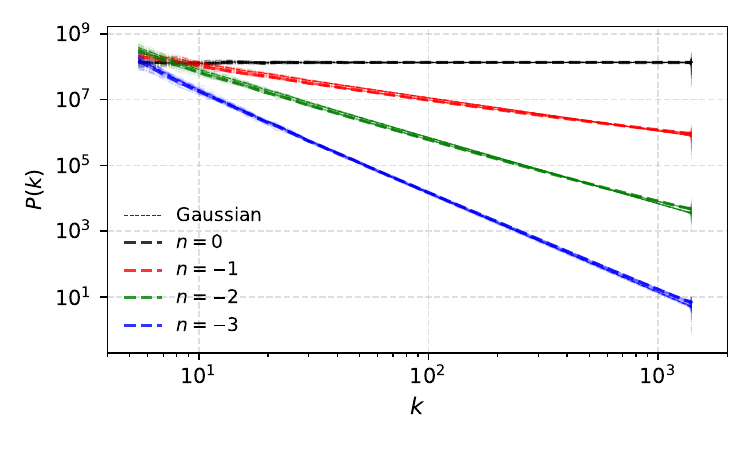}
    \vs{-3mm}
    \caption{
        Power-law power spectra given by Eq.~\eqref{eq:Power-Spectrum} with spectral indices $n=0,-1,-2,-3$ are shown for the Gaussian random field (dotted lines) and for the corresponding exponential random field (dash-dotted lines) obtained via the mapping in Eq.~\eqref{eq:Gauss2ExpTransfo}. For each value of $n$, we plotted $20$ individual realizations of the Gaussian field and the power spectra of their corresponding mapped exponential random field. The less transparent curve for each $n$ denotes the power spectrum averaged over realizations. Notice that, for all considered $n$, the Gaussian and exponential power spectra curves entirely overlap on top of each other, both at the level of individual realizations and therefore also in their average. This demonstrates that the transformation in Eq.~\eqref{eq:Gauss2ExpTransfo} preserves the power-law spectrum for $n = 0,\.-1,\.-2,\.-3$ to a high accuracy.\vs{-3mm}}
	\label{fig:power_spectra} 
\end{figure}

Following the generation of $F$, we proceed to map it to an exponential random field $E$ in two steps. First, we map $F( \mbm )$ into a uniform distribution $U( \mbm ) \in [ 0,\.1 ]$ by taking its cumulative distribution function (CDF). Second, we map $U( \mbm )$ to an exponential random field $E( \mbm )$ using the inverse CDF of the exponential distribution. This process defines the mapping $F( \mbm ) \rightarrow E( \mbm )$, with
\vs{2mm}
\begin{align} \label{eq:Gauss2ExpTransfo}
	E( \mbm )
		= \!
			- \.
			\frac{ 1 }{ \lambda }\!
			\ln\!
			\left\{
				1
				-
				\frac{ 1 }{ 2 }\!
				\left[
					1
					+
					\text{erf}\mspace{-2mu}
					\left(
						\frac{ F( \mbm ) - \mu[ F ]}
						{ \sqrt{2}\.\zeta[ F ] }
					\right)
				\right]
			\right\}
			,
            \\[-4mm]
            \notag
\end{align}
where $\mu[ F ]$ and $\zeta[ F ]$ are the mean and standard deviation of $F$, respectively. This mapping aims to preserve the power spectrum and its spatial correlation, where the preservation of the power spectrum can be seen from the overlapping curves in Figure~\ref{fig:power_spectra}. Note that, on one hand, this transformation is purely technical and is not intended to represent any physical process responsible for generating non-Gau{\ss}ianity. On the other hand, the exponential random field $E( \mbm ) \in [ 0,\.\infty )$ should be considered as an effective model for studying the extreme value statistics of a random field that possesses an exponential tail rather than a complete perturbation distribution.
% Nevertheless, this mapping is completely general and able to accommodate a larger set of power spectra beyond that in Eq.~\eqref{eq:Power-Spectrum}, as long as the induced spatial correlation has a correlation length $l_{\crm} \ll L$ for the underlying $F( \mbm )$ that defines $E( \mbm )$. 
In the following, for brevity, we will use $\mu = 0$ and $\zeta = \lambda = 1$ without loss of generality in the context of extreme value statistics, since our analysis concerns the probability distribution as a function of rarity, which is invariant under rescaling of $\lambda$.
\vs{3mm}

%%%%%%%%%%%%%%%%%%%%%%%%%%%%%%%%%%%%%
\noindent
{\it Extreme-Value Statistics\,---}\vp\;In many cases, one is interested in an average of a number of random variables. There are, however, many situations in which the relevant quantities are extreme fluctuations, such as {\it block maxima}:
\vs{-1mm}
\begin{align} 
	M_{B}
		\equiv
			\max_{i\mspace{1mu}=\mspace{1mu}1,
			\mspace{1mu}\ldots,\mspace{1mu}B
			\vphantom{\big]}} \Delta_{i}
			\, ,
\end{align}
with $B \in \Nbb$ independently and identically-distributed (iid) random variables $\{ \Delta_{i} \}_{i\mspace{2mu}=\mspace{2mu}1}^{B}$, where in our work, the $\Delta_{i}$ correspond to the realisations of the random field $F$ or $E$ on the set of pixels within a given block [e.g.~constituted by a number of Hubble patches in the case of PBH formation (see below)], sampled over a sufficiently large number of blocks. According to the Fisher--Tippett theorem~\cite{fisher-tippett-1928}, the distribution of $M_{B}$ converges to one of the {\it generalised extreme value} (GEV) distributions, which are inherently non-Gau{\ss}ian (see below) \footnote{It is important to stress the non-perturbative nature of the non-Gau{\ss}ianity in this work. In cosmology, the commonly used parameters like $f_{\rm NL}$ or $g_{\rm NL}$ describe an expansion around Gau{\ss}ianity. While it might well be that for a given rarity, such as $6\,\sigma$, the mentioned expansion might actually be sufficient for the derivation of the studied effects, the exact exponential tail, however, cannot be obtained from any finite expansion based on Gau{\ss}ianity.}.
\vs{5mm}
\newpage

Specifically, if there exist two sequences $\{ a_{i} > 0 \}_{i\mspace{2mu}=\mspace{2mu}1}^{\infty}$ and $\{ c_{i} \in \Rbb \}_{i\mspace{2mu}=\mspace{2mu}1}^{\infty}$ such that
\begin{align} 
	\lim_{ i \to \infty } {\rm Prob}
	\bigg[
		\frac{ \Delta_{i} - a_{i} }
		{ c_{i} }
			<
				\delta
	\bigg]
		=
			H( \delta )
			\, ,
\end{align}
then $H( \delta )$ {\it inevitably} must be one of the GEV distributions (cf.~Ref.~\cite{10.5555/262578}):
\begin{align}
	- \log\mspace{-2mu}
	\big[
		H^{s}_{\alpha,\.\gamma}( \delta )
	\big]
		&=
			\begin{cases}
				\frac{ 1 }{ \gamma }\!
				\left[
					1
					+
					s\!
					\left(
						\frac{ \delta\mspace{1.5mu}
						-\mspace{1.5mu}\alpha }
						{ \gamma }
					\right)
				\right]_{}^{-1 / s }
				 &
					( s \neq 0 )
					\\[3mm]
			\frac{ 1 }{ \gamma }
			\exp{\!
			\left[
				\!
				\left(
					\frac{ \delta\mspace{1.5mu}
					-\mspace{1.5mu}\alpha }
					{ \gamma }
				\right)
			\right]}
			\vphantom{\bigg]}
				 &
					( s = 0 )
		\end{cases}
		.
		\label{eq:P-Generalised-EVDs}
\end{align}

\noindent The associated probability density function (PDF), $h^{s}_{\alpha,\mspace{1mu}\gamma}$, is related to the CDF $H^{s}_{\gamma,\mspace{1mu}\alpha}$ via
\begin{align}
	H^{s}_{\alpha,\mspace{1mu}\gamma}( \delta )
		&=
			\int_{- \infty}^{\delta}\d y\;
			h^{s}_{\alpha,\mspace{1mu}\gamma}( y )
			\, .
			\label{eq:PDF-Generalised-EVD-integral}
\end{align}
Here, $s$, $\alpha$, and $\gamma$ are the {\it shape}, {\it location}, and {\it scale parameters}, respectively, to be determined from data. The choices $s = 0$, $s > 0$, and $s < 0$ correspond to the {\it Gumbel}, {\it Fr{\'e}chet}, and {\it Weibull} distributions. Any distribution of iid random variables for which the limit exists, converging to one of those three distributions, belongs to their {\it maximum domain of attraction} (MDA) (cf.~Ref.~\cite{10.5555/262578}). For $n = 0$, the Gau{\ss}ian and the exponential distribution belong to the MDA of the Gumbel distribution.
\vs{5mm}
\newpage

Of course, for our studies with non-zero spectral index, the random variables are not truly independent, hence violating assumptions of the Fisher--Tippett theorem. However, as we will see below, the tails of the resulting block-maxima distributions will nevertheless follow a Gumbel distribution.

%%%%%%%%%%%%%%%%%%%%%%%%%%%%%%%%%%%%%
\noindent
{\it Correlation Length\,---}\vp\;When measuring the block maxima of spatially-correlated random fields, the characteristic length scale of spatial correlation $l_{0}$ is necessary for selecting an appropriate box size. In this context, we refer to $l_{0}$ as the \textit{correlation length}, being defined as (cf.~Ref.~\cite{Naselsky:1998by, Kasak:2020vtd})
\begin{align}
\label{eq:corr-length}
	l_{0}
		\equiv
			\frac{ \zeta_{0} }{ \zeta_{1} }
			\, ,
\end{align}
where $\zeta_{0}$ and $\zeta_{1}$ are the standard deviations of the random fields $F( \xbm )$ or $E( \xbm )$ and of their gradient, respectively. Intuitively, $l_{0} = \langle k^{2} \rangle_{P}^{-1/2}$ estimates spatial correlations by the weighted average of $k^{2}$ by $P( k )$ for continuous random fields. In numerical realisations, spatial derivatives are replaced by the central finite differences, such that $l_{0} = \big\langle \sin^{2}( \kbm \cdot \Delta x_{i} )/{\Delta x_{i}}^{2} \big\rangle_{P}^{-1/2} $, for $i = 1,\.2,\.3$. This gives $l_{0} \sim \langle k^{2} \rangle_{P}^{-1/2}$ for small $k$, while larger $k$ (finer structures) are suppressed due to pixelisation.

\begin{figure}[t!] 
	\centering
	\includegraphics[width=0.90\columnwidth]{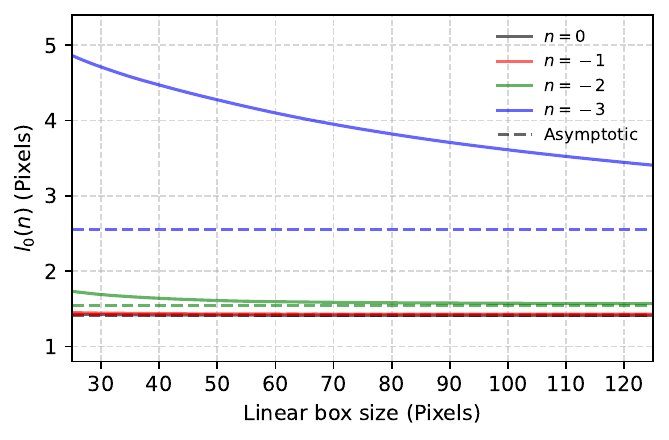}
	\caption{
		Correlation length~\eqref{eq:corr-length} of exponential random fields for different spectral indices, with a rare event of $6\.\sigma$ at the box centre. The flat dotted horizontal lines indicate the respective asymptotic correlation length derived from the full random field, while the solid lines depict the correlation length as a (diminishing) function of the box size. Note that the curves for $n = 0$ and $n = -1$ largely overlap.
		\vs{-2mm}
	}
	\label{fig:correlation-length} 
\end{figure}

Figure~\ref{fig:correlation-length} shows the numerical estimation of $l_{0}$ of exponential random fields with different spectral indices, guaranteed with a $6\.\sigma$ event at the box centre. The flat line represents the asymptotic value, calculated using Eq.~\eqref{eq:corr-length} by sampling over $10^{4}$ exponential random fields. Decay of the curves with increasing block size can be observed, the smaller values of $n$ yield a slowed decline.

In the following, we will consider a cubic block with linear size of $33$ pixels. For $n = -3$, this is approximately $6$ times the correlation length inside the block, and $12$ times the asymptotic value (see Fig.~\ref{fig:correlation-length}). At this size, the curves for $n = 0,\.-1$ have already reached the asymptotic value, but not for $n = -2,\.-3$. This indicates that the chosen block size can capture substantial correlation for $n = -2,\.-3$, but is too large for $n = -1$ to detect any correlation effect.

\begin{figure}[t!] 
	\centering
    \vs{-1mm}\includegraphics[width=0.95\columnwidth]{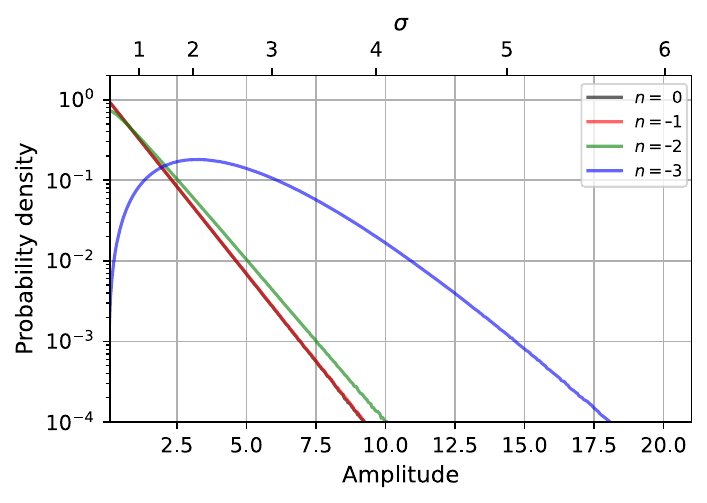}
	\caption{
		Probability density distribution of exponential random fields {\it within} a block of $33$ pixels (linear) dimension for different spectral indices (see legend). Note again the overlap of the curves for $n = 0$ and $n = -1$.
	}
	\label{fig:nn-population} 
\end{figure}

The effect of spatial correlation is shown in Fig.~\ref{fig:nn-population}, which shows the PDFs of the exponential random field within a block size of $33$ pixels, and containing at least one extreme-value event with at least $6\.\sigma$ rarity. For $n = -3$, we observe a particularly strong suppression of low-amplitude events in the neighbourhood compared to the uncorrelated case ($n = 0$).

\begin{figure}[t!] 
	\centering
	\vs{1mm}\includegraphics[width=0.95\columnwidth]{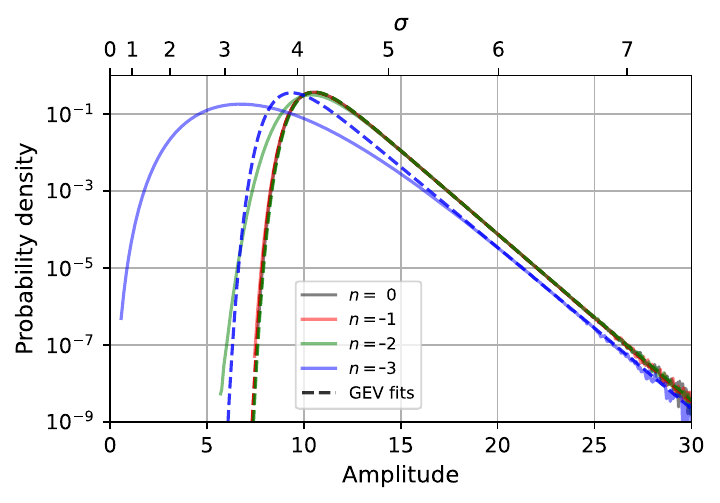}
	\caption{
		Block-maxima PDF obtained by sampling exponential random fields of size $2^{9}\mspace{-2mu}\times 2^{9}\mspace{-2mu}\times 2^{9}$ pixels for each spectral index with rarity up to and beyond $7.5$ sigma ($1$ in $10^{13}$). The dashed line shows the corresponding best-fit GEV distribution [$\mspace{1mu}$cf.~Eq.~\eqref{eq:P-Generalised-EVDs}] between amplitude values $20$ and $30$. 
		The parameters are
		  $( \alpha,\.\gamma,\.s ) = 
		  	( 10.52 ,\.1.00 ,\.-9.84\times 10^{-5} \simeq 0 )\;
			[\mspace{1mu}n = 0\mspace{1mu}]$,
		  $( 10.51 ,\.1.00,\. -3.27\times 10^{-4} \simeq 0 )\;
		  	[\mspace{1mu}n = -1\mspace{1mu}]$, 
			$( 10.54 ,\. 0.99 ,\. 9.86\times 10^{-4} \simeq 0 )\;
			[\mspace{1mu}n = -2\mspace{1mu}]$, 
		and 
		  $( 9.36 ,\.1.04 ,\. 4.50\times 10^{-4} \simeq 0 )\;
		  [\mspace{1mu}n = -3\mspace{1mu}]$. 
		As before, the curves for $n = 0$, $-1$ essentially overlap.
	}
	\label{fig:block-maxima-GEV}
\end{figure}

%%%%%%%%%%%%%%%%%%%%%%%%%%%%%%%%%%%%%
\noindent
{\it Spatial-Correlated Random Fields\,---}\vp\;The PDF of block maxima is sensitive to spatial correlation, which is shown in Fig.~\ref{fig:block-maxima-GEV}. The ($n = 0$)-curve represents the PDF of block maxima sampled from an uncorrelated exponential field, which must follow a Gumbel distribution~(cf.~Ref.~\cite{10.5555/262578}). Any deviation from the case $n = 0$ in Fig.~\ref{fig:block-maxima-GEV} is solely due to the effect of spatial correlation in the random field, as the spectral indices are the only varying parameter. The more the correlation length (within a block) deviates from its asymptotic value (cf.~Fig.~\ref{fig:correlation-length}), the more the PDF of block maxima deviates from the case $n = 0$.

The key result of Fig.~\ref{fig:block-maxima-GEV} is that spatial correlations tend to broaden the PDF of the block maxima, by enhancing the probability of finding low-amplitude block maxima while suppressing high-amplitude ones. This suppression occurs because spatial correlation causes rare events of extreme values to cluster, as can be observed in Fig.~\ref{fig:nn-population}. However, block maxima only records the strongest signal within a given block, while the rest are discarded. This is in contrast with uncorrelated random fields ($n = 0$), where rare events are evenly spread, being hence more likely to be sampled as block maxima.

\begin{figure*}[t!] 
	\centering
	\vs{-2mm}
	\includegraphics[width=0.92\linewidth]{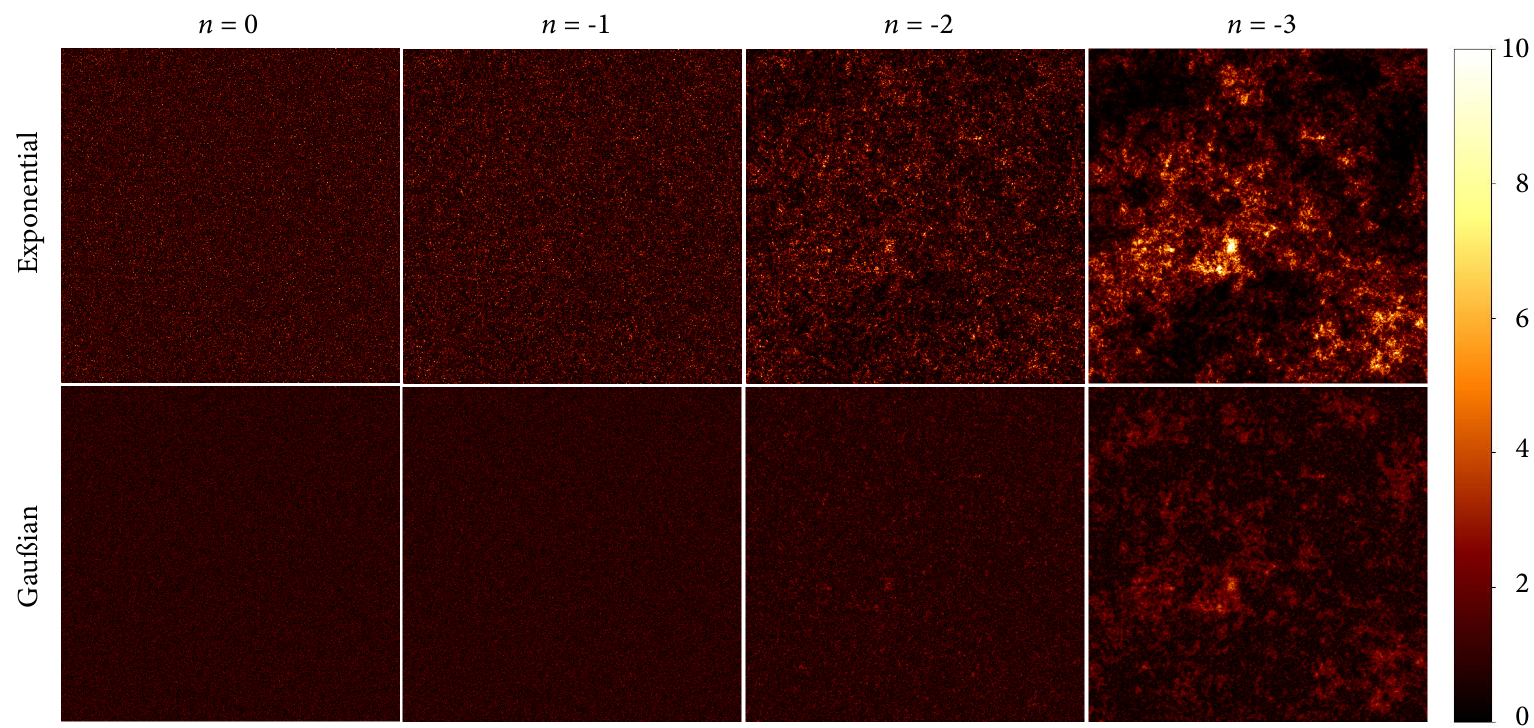}
	\caption{
		Two-dimensional slices from a three-dimensional random field simulation, with the top row showing exponential random fields and the bottom row indicating Gau{\ss}ian random fields, all cases have unit variance. The images, from left to right, display increasing spatial correlation in the random fields for spectral indices $n = 0,\.-1,\.-2,\.-3$, respectively. The colour bar indicates the amplitude, with the brightest signals having a value of at least $10$.
        \vs{-2mm}
		}
	\label{fig:intuitive} 
\end{figure*}

Similarly, for spatially-correlated random fields, low-amplitude events tend to cluster and form large under-dense regions, as can be well observed in Fig.~\ref{fig:intuitive}, most pronounced for $n = -3$. Therefore, these under-dense regions increase the likelihood of finding low-amplitude events as block-maxima. While spatial correlation broadens the PDF of block maxima, beyond the $6$-sigma level, their tails show similar statistical properties across all spectral indices, exhibiting exponential decays at the same rate, shown as parallel slopes on the logarithmic scale in Fig.~\ref{fig:block-maxima-GEV}. This behaviour arises because such extreme events are exceedingly rare, occurring at most once per simulation, or are rare enough to materialise only in one block. Accordingly, despite the presence of spatial correlation, rare events beyond the $6$-sigma level are recorded as block maxima in a statistically-independent manner and are well-described by the tail of a Gumbel distribution ($s = 0$), as they are sampled independently from different exponential field simulation. This shows that $\alpha$ is the only parameter sensitive to changes in $n$ for modeling tails behaviors in Fig.~\ref{fig:block-maxima-GEV}. Despite not being in the main focus of our studies, we note that also the block size influences the fitting parameters, with diminishing effect when the former increases.

\begin{figure}[t!] 
	\centering
	\vs{2mm}\includegraphics[width=0.9\columnwidth]{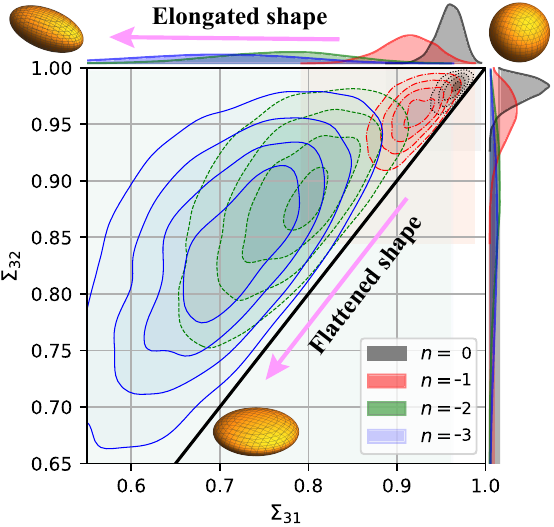}
    \vs{2mm}\includegraphics[width=0.9\columnwidth]{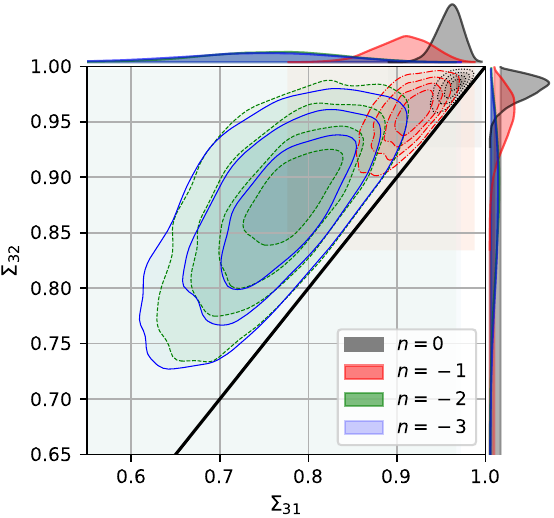}
	\caption{
		Ratio of weighted principal components $\Sigma_{32}$ against $\Sigma_{31}$ obtained from sampling at least $10^{5}$ simulations that contain at least one extreme-value event of $6$ sigma rarity. The black straight line marks $\Sigma_{31} = \Sigma_{32}$. Equal probability contours are shown, representing the weighted principal components for a cluster with rare fluctuation (of $6$ sigma and beyond) in its centre. The upper panel shows the results for exponential random fields, while the lower panel depicts the Gau{\ss}ian case.
		}
	\label{fig:sphericity} 
\end{figure}

%%%%%%%%%%%%%%%%%%%%%%%%%%%%%%%%%%%%%
\noindent
{\it Clustering\,---}\vp\;Spatial clustering of large peaks \footnote{We appreciate that most of the literature discusses continuous situations in which a {\it peak} is an extended feature (unless a delta-peak is assumed). On the contrary, in our discrete study we refer to a peak as an excess in a single pixel (representing an initial Hubble patch at the time of horizon crossing of an overdensity).} generated by random fields plays important r{\^o}les in many mathematical and physical contexts (cf.~Ref.~\cite{Bardeen:1985tr}). Depending on the specific type of the random field, in particular its spatial correlations, the clusters can emerge of various sizes and shapes. We will see that the spectral index [cf.~Eq.~\eqref{eq:Power-Spectrum}] plays an important r{\^o}le for these characteristics. Below we apply this to PBH formation and demonstrate the potential large impact of non-white-noise cases.

We have performed a cluster-shape analysis. For a pronounced demonstration of the resulting effects and to connect to subsequent analyses, we focus on clusters containing at least one $6$-sigma signal in their centres. For our analysis, we will make use of the {\it Weighted Principal Component Analysis} (WPCA)~\cite{10.1093/mnras/stu2219}, which allows us to quantify the sphericity in the random-field clusters in terms of distribution variance. A spherically-distributed random field must exhibit equal variance along all principal components, whereas non-sphericity manifests as differences in them. 

In contrast to the conventional principal component analysis (PCA)~\cite{doi:10.1080/14786440109462720, Hotelling:1933aaa}, which treats all data equally, WPCA discriminates individual data points and features by assigning weights to them. Specifically, WPCA minimises
\vs{-1mm}
\begin{align} \label{eq:WPCA}
	\chi^{2}
		=
			\sum_{i,\mspace{1mu}j}\.W_{ij}
			\left(
				X_{ij} - \sum_{k} P_{ik}\.C_{kj}
			\right)^{\!2}
			\; .
\end{align}
Here, $\Xbm \equiv [ \xbm_{1}, \ldots, \xbm_{B} ]$ is the grid matrix, with positions $\xbm_{i} \equiv [\mspace{1mu}x_{i}, y_{i}, z_{i}\mspace{1mu}]{}^{\!\Tsf}$, where $i \in [\mspace{1mu}0,\.B\mspace{1mu}]$ for {$i,\.B \in \Nbb$}, with $B$ being the total number of grid points in the random-field block. For our purpose, each spatial coordinate must be weighted equally by the field amplitude at a given position, such that $\Wbm \equiv \big[\mspace{1mu}F( \xbm_{1} )\,\Ibb,\.\ldots,F( \xbm_{B} )\,\Ibb\.\big] / \sum_{i} F( \xbm_{i} )$, where $\Ibb \equiv [\mspace{1mu}1,\.1,\.1\mspace{1mu}]_{}^{\Tsf}$. The quantity $\chi^{2}$ is then minimised by fitting the coefficient matrix $\Cbm$ with the principal-components matrix $\Pbm$, whose rows and columns correspond to spatial coordinates and components, respectively; the rows of $\Cbm$ correspond to components and its columns to grid points.

The principal components can be diagonalised into the distribution-variance matrix $\Sigma \equiv \Cbm\mspace{1mu}\Cbm^{\Tsf}$, with $\Sigma_{i} \geq \Sigma_{i+1}$ $\forall\;i \in \Nbb$. These variances measure the sphericity of the random field within a block by measuring its anisotropy. The ratio $\Sigma_{31} \equiv \Sigma_{3} / \Sigma_{1} \in [\mspace{1mu}0,\.1\mspace{1mu}]$ reflects isotropy when $\Sigma_{31} = 1$, and anisotropy for $\Sigma_{31} < 1$. If clusters form due to spatial correlation, $\Sigma_{31}$ indicates whether the cluster is spherically symmetric. In the presence of anisotropy, the cluster shape can be characterised by the ratio $\Sigma_{32} \equiv \Sigma_{3} / \Sigma_{2} \in [0, 1]$: $\Sigma_{32} = 1$ implies an ellipsoidally-shaped cluster, $\Sigma_{32} \sim \Sigma_{31}$ (black line) indicates a circular disk shape; $\Sigma_{32} < 1$ corresponds to a disk-like shape.

Figure~\ref{fig:sphericity} illustrates the ratios of principal components characterising the geometric properties of the random field. It is clear that non-sphericity develops as spatial correlation increases. In the absence of spatial correlation ($n = 0$), both $\Sigma_{31}$ and $\Sigma_{32}$ have mean values closer to $1$ of minor standard deviation, indicating the random field is mostly homogeneous and isotropic. As the spectral index increases, particularly at $n = -3$, the mean of $\Sigma_{31}$ and $\Sigma_{32}$ decrease, indicating that the formed cluster is preferably anisotropic. Furthermore, as the standard deviation increases, it indicates a large variety in the cluster shapes, either forming elongated or flattened shapes.

To isolate the effect of the tail contribution from that of the spatial correlations, we have also performed simulations of Gau{\ss}ian random fields with the same set of power spectra as for exponential fields (see the two panels of Fig.~\ref{fig:sphericity} for a direct comparison). Our results show that non-Gau{\ss}ianity amplifies the degree of non-sphericity generated by spatial correlation, and produces a broader variety of cluster shapes. For $n = -2$ and $n = -3$, where large spatial clusters are formed, the exponential tails in the exponential random field enhance the degree of spatial anisotropies, even though these Gau{\ss}ian and exponential random fields shares identical power spectra by construction of Eq.~\eqref{eq:Gauss2ExpTransfo}. This occurs because the non-sphericity measure~\eqref{eq:WPCA} weights the amplitudes of the random field within each block. 
Hence, in the exponential case, the existence of a heavy tail leads to a greater variance of amplitude within a cluster; therefore, the presence of an at least $6\.\sigma$ event within a cluster can change $\Sigma_{ii}$, for $i = 1,\,2,\,3$, more substantially when compared to the Gau{\ss}ian case.

%%%%%%%%%%%%%%%%%%%%%%%%%%%%%%%%%%%%%
\noindent
{\it Application: Primordial Black Holes\,---}\vp\;Primordial black holes~\cite{ZeldovichNovikov69, Carr:1974nx} form in the early Universe, usually before matter-radiation equality and have long been discussed as dark matter candidates~\cite{1975Natur.253..251C}; see Ref.~\cite{Carr:2020xqk} for a recent review. After the detection of gravitational waves from black hole binaries~\cite{TheLIGOScientific:2016pea}, it was realised these could be primordial in Nature~\cite{Bird:2016dcv}.

There are numerous formation scenarios (see Ref.~\cite{ESCRIVA2024261} for the currently most extensive review), but all have in common that they yield overdensities which are sufficiently large to collapse to a black hole. This could have happened for Hubble-horizon-sized overdensities with density contrast $\delta$ exceeding a certain critical threshold $\delta_{\crm}$, which depends on the shape of the overdensities and their statistics, the collapsing medium (with particular sensitivity to the equation of state~\cite{Carr:2019kxo}) as well as the presence of nearby overdensities~\cite{Escriva:2023qnq}. For an isolated density peak, assuming spherical symmetry and collapse during radiation domination, simulations give $\delta_{\crm} = 0.45$~\cite{Musco:2012au}\footnote{More advanced estimates utilise the so-called compaction function (see, e.g., Ref.~\cite{Shibata:1999zs}){\,---\,}an approach which unavoidably induces some degree of smoothing, hence{\,---\,}given the discrete (pixelated) approach used in this work{\,---\,}it is a relevant question to ask whether and how the use of the smoothing due to the use of compaction functions alters the results. This could certainly have large implications on subhorizon scales, but in our work, each pixel is thought to represent an entire initial Hubble patch, so we do not expect a strong effect.}. Since this value is about four orders of magnitude larger than that of typical CMB fluctuations~\cite{Planck:2018jri}, PBH formation fundamentally depends on the tail of the density distribution, making it inherently rare, usually beyond five sigma. Importantly, the collapse is not instantaneous, typically taking several e-folds (see Refs.~\cite{Musco:2008hv, Musco:2012au, EscrivaManas:2021mqb}). Especially if other large nearby overdensities are present, it can take around $10^3$ Hubble times until the final black hole mass is reached~\cite{Escriva:2023qnq}.

Here, we will study exactly the effect of finite formation time. This implies the consideration of blocks constituted by the multiple initial Hubble volumina (from the time of horizon crossing) contained in the Hubble patch at the later time of PBH formation. The nature of gravitational collapse implies that the largest of the involved overdensities will be key for the final black hole. Hence, this situation naturally leads to considering block maxima. It has been shown that the probability distributions in many cases have exponential tails (cf.~Refs.~\cite{Biagetti:2018pjj, Ezquiaga:2019ftu, Tada:2021zzj}), these being in our main focus.

At present, there exists no measurement of the primordial power spectrum on scales $k_{\rm PBH}$ relevant for PBH formation. It is clear, however, that the near-scale-invariant form determined at CMB scales cannot hold around $k_{\rm PBH}$, but must rather dramatically increase to yield enough sufficiently large overdensities undergoing gravitational collapse. Due to the mentioned uncertainty, one has to make assumptions, and here we will assume that for a range of scales relevant for PBH formation, the primordial power spectrum can be approximated by the power-law form of Eq.~\eqref{eq:Power-Spectrum}. This is somewhat arbitrary, but our results will qualitatively hold whenever the slope of the power spectrum is similar.

%As can be observed in Fig.~\ref{fig:block-maxima-GEV}, the effect of spatial correlation tends to suppress the total number of PBHs, at the cost of enhancing the total number of voids that can be found in a given Hubble patch. While the population is reduced, the total mass involved in the gravitational collapse increases with the spatial correlation length, which may already be inferred from Fig.~\ref{fig:nn-population}.

Due to the partly strong clustering, especially in the case of $n = -3$, it is important to consider the effect of neighbouring overdensities on the PBH formation. Conveniently, Ref.~\cite{Escriva:2023qnq} has recently performed a detailed numerical study on how the presence of an additional overdensity alters the black hole formation process. 
%Their numerical results allow for several important conclusions: The presence of an additional overdensity can {\it 1.)} lower the PBH formation threshold, {\it 2.)} increase the final black hole mass, and {\it 3.)} prolong the time until mass saturation is reached. Furthermore, for these effects to happen, both overdensities have to practically be directly neighbouring; for larger separations the collapses proceed independently.
For our analysis, we extract the final PBH mass, as a function of horizon mass, from the numerical results in Tab.~1 of the mentioned paper. Together with its Fig.~9, and subsequent interpolation, we obtain an approximation of the PBH mass function resulting from two overlapping overdensities. At present, no extension of the work~\cite{Escriva:2023qnq} %, which uses state-of-the-art numerical simulations to study gravitational collapse of super-horizon curvature fluctuations, 
exists to go beyond two overlapping overdensities, which would be needed in our strongly-clustered cases. We will hence na{\"i}vly assume that, to lowest order, the individual effects of each additional overdensity are additive, knowing that such situations can only be properly addressed through (yet to be performed) numerical simulations. Hence, our results{\,---\,}despite being qualitatively correct{\,---\,}should quantitatively be taken {\it cum grano salis}. On the one hand, the assumed additivity may overestimate our results; on the other, we have conservatively neglected two effects which clearly underestimate the strength of our results: firstly, the presence of non-nearest neighbours, and, secondly, the accretion phase which Escriv{\`a}--Yoo~\cite{Escriva:2023qnq} show to significantly increase the black hole mass, this being possibly enhanced by the clustering of overdensities. A more precise evaluation is needed for a full quantitative understanding of this new effect. 
%Here, we point out these novel effects for the first time.}

\begin{figure}[t!] 
	\centering
	\vs{2mm}\includegraphics[width=0.95\columnwidth]{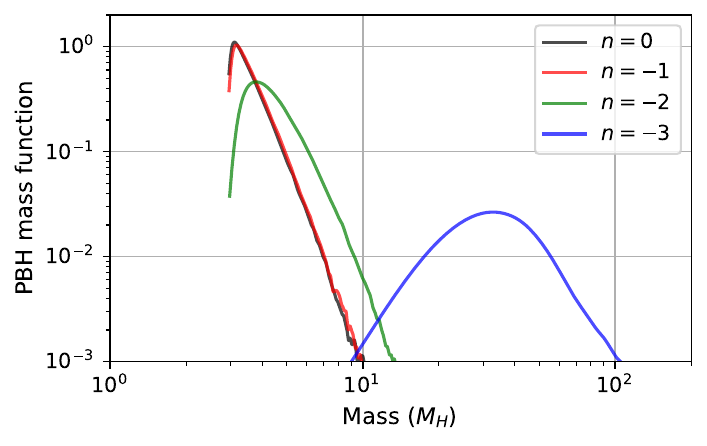}
	\caption{
		Probability density distribution of the PBH mass (in unit of horizon mass $M_{\mspace{-1mu}H}$) obtained from sampling at least $10^{5}$ exponential field simulations that contain at least one extreme-value event of $6$ sigma rarity. The mass spectrum considers gravitational collapse of the beyond-$6$-sigma fluctuations, taking neighbouring overdensities into account.
        \vs{2mm}
		}
	\label{fig:neighbours} 
\end{figure}

Figure~\ref{fig:neighbours} shows our findings for the PBH mass functions{\,---\,}as before, for the four cases of white noise ($n = 0$) and non-zero spatial correlations ($n = -1,\.-2,\.-3$). As mentioned, as a rough approximation, we have assumed that the effects on threshold lowering and collapse-time prolongation, leading to an overlap-induced increase of the PBH mass, are additive for additional neighbouring overdensities. It can be observed that the latter can have a large effect in various ways, yielding a shift, broadening\footnote{Observe that the broadening of the PBH mass function does not only extend towards higher masses, due to the absorption of clustered peaks, but also towards lower masses, since neighbouring (partly overlapping) peaks generically lower the PBH formation threshold (see Ref.~\cite{Escriva:2023qnq}).} and lowering of the PBH mass function whose minimum mass increases with increasing spatial correlation. This is due to the latter enabling the process of absorption of nearest neighbouring peaks. Of course, the exact numerical values remain to be obtained from a detailed simulation, but one can be certain to expect the effect to remain and to be at least of similar magnitude. There are, of course, several other effects which alter the shape of the mass function, such as a softening of the equation of state (cf.~Ref.~\cite{Carr:2019kxo}). While most of these alter the number of PBHs at a certain mass, the novel effect described in this work additionally implies a substantial shift towards larger PBH masses.\footnote{Note that this effect cannot directly compare to the Gau{\ss}ian random field case, since Gau{\ss}ian random fields do not possess a heavy tail. Specifically, if the threshold for PBH formation is set to $20$ standard deviation ($6 \sigma$ rarity) of the exponential random field, then a Gau{\ss}ian random field of the same variance would yield essentially no threshold exceedance.}

Depending on the value of the spectral index, even two above-threshold peaks can occur right next to each other. Concretely, for six-sigma fluctuations, the corresponding probabilities are $10^{-9}$ for $n = 0$, $10^{-7}$ for $n = -1$, $10^{-4}$ for $n = -2$, and $10^{-2}$ for $n = -3$. The associated directly-neighbouring overdensities will clearly lead to the formation of a single black hole instead of separate ones. Similarly, also slightly more distant overdensities can be expected to coalesce. %However, substantiating the latter statement requires dynamical calculations which are beyond the scope of this article. In any case, the mentioned effects will suppress the number of PBHs with a corresponding mass increase. We will come back to this important topic in a future publication.
Besides nearest neighbours, in a strongly-clustered situation, also more distant high-amplitude fluctuations can be assumed to be relevant. In general, the cluster-induced non-sphericities (cf.~Fig.~\ref{fig:sphericity}) can be expected to substantially alter the initial PBH spin distribution, leading to larger initial spins than previously expected (cf.~Ref.~\cite{DeLuca:2020jug} for prior work studying the spin distribution of clustered PBH populations).

Furthermore, a fraction of separate, but sufficiently near-by, PBHs can be expected to undergo merger processes, thereby implying additional changes in spin and mass. These effects will be left for future work, which will also include the study of more realistic primordial power spectra. Indeed, PBH formation without overproduction cannot originate from pure power-law power spectra studied in this work, but must involve one (or several) peak(s). Any peak has a rising and a falling side, for which power-law power spectra may serve as a first approximation. While this might be very rough, the purpose of our work is to show as a proof of principle that an important class of power spectra will have a strong effect on clustering and sphericities as well as the PBH mass function. These novel effects will certainly be present for any peaked primordial power spectrum which leads to PBH formation. Finally, we would like to point out that, since all studied effects predominantly involve overdensities deep in the distribution of the tail, they would all be affected similarly by an overall normalisation.

\noindent
% {\it Conclusion\,---}\vp\;In this {\it Letter}, we have studied large-scale spatially-correlated exponential random fields, particularly investigating clustering and non-sphericity, which we found to be strongly affected by the spatial correlation. 
{\it Conclusion\,---}\vp\;In this {\it Letter}, we have studied large-scale spatially-correlated exponential random fields, particularly investigating the rare-event statistics of clustering and non-sphericity developed by the spatial correlation and the existence of an exponential tail. For our studies, we performed the currently largest corresponding simulation, permitting to resolve beyond-$7.5$-sigma events ($1$ in $10^{13}$). In the context of extreme-value theory, we have shown that rare events in spatially-correlated exponential random fields can be modelled using generalised extreme-value distributions.

We have further demonstrated that, for sufficiently rare events, this distribution reduces to the same family of extreme-value distributions as in their independent case for rare events, i.e.~the Gumbel distribution. More generally, if a sequence of random events is large compared to the correlation scale of the random variable, then it is possible to extend the application of Gumbel distribution to describe random variables that are identical but not independently-distributed even in non-asymptotic models.

Therefore, as an application, we were able to study individual Hubble patches which would fulfil the condition for the formation of primordial black holes. Here, we point out a novel effect manifesting in sizeable spatial-correlation-induced shift, broadening and lowering of their mass spectrum. The emerging non-sphericity will likely also alter their initial spin distribution, which we will investigate in a forthcoming publication. Extending the applications of extreme-value distributions to other cosmological situations in which spatially-correlated random fields are relevant may reveal other novel effects and will provide further insights into the statistical nature of rare events in our Universe.

\vp
{\it We thank the anonymous referee for useful input.} DJS acknowledges support of Deutsche Forschungsgemeinschaft (DFG, German Research Foundation) -- project number 315477589 -- TRR 211. KHC acknowledges the supports of the Excellence Cluster ORIGINS. The authors gratefully acknowledge the {\it Gau{\ss} Centre for Supercomputing e.V.}~(\href{www.gauss-centre.eu}{www.gauss-centre.eu}) for funding this project by providing computing time on the {\it GCS Supercomputer SuperMUC at Leibniz Supercomputing Centre} (\href{www.lrz.de}{www.lrz.de}).

\vfil
%\newpage

%%%%%%%%%%%%%%%%%%%%%%%%%%%%%%%%%%%%%
\setlength{\bibsep}{4pt}
\bibliography{references}

\end{document}